\documentclass[twocolumn,english,amsart,showpacs,preprintnumbers,amsmath,amssymb,floatfix]{revtex4-1}
\usepackage{tikz,xcolor}
\usepackage[colorlinks = true,
linkcolor = blue,
urlcolor  = blue,
citecolor = blue,
anchorcolor = blue]{hyperref}
\RequirePackage{graphicx}
\definecolor{lime}{HTML}{A6CE39}
\DeclareRobustCommand{\orcidicon}{%
	\begin{tikzpicture}
		\draw[lime, fill=lime] (0,0)
		circle [radius=0.16]
		node[white] {{\fontfamily{qag}\selectfont \tiny ID}};
		\draw[white, fill=white] (-0.0625,0.095)
		circle [radius=0.007];
	\end{tikzpicture}
	\hspace{-2mm}
}
\foreach \x in {A, ..., Z}{%
	\expandafter\xdef\csname orcid\x\endcsname{\noexpand\href{https://orcid.org/\csname orcidauthor\x\endcsname}{\noexpand\orcidicon}}
}

\usepackage[T1]{fontenc}
\usepackage[latin9]{inputenc}
\usepackage{color}
\usepackage{array}
\usepackage{amstext}
\usepackage{graphicx}
\usepackage{esint}
\usepackage{rotating}
\usepackage[font=small,labelfont=bf]{caption}
\usepackage{xcolor}
\usepackage{multirow}
\usepackage{booktabs}
\makeatletter

\@ifundefined{textcolor}{}
{%
 \definecolor{BLACK}{gray}{0}
 \definecolor{WHITE}{gray}{1}
 \definecolor{RED}{rgb}{1,0,0}
 \definecolor{GREEN}{rgb}{0,1,0}
 \definecolor{BLUE}{rgb}{0,0,1}
 \definecolor{CYAN}{cmyk}{1,0,0,0}
 \definecolor{MAGENTA}{cmyk}{0,1,0,0}
 \definecolor{YELLOW}{cmyk}{0,0,1,0}
 }

\@ifundefined{definecolor}
 {\usepackage{color}}{}
\@ifundefined{definecolor}
 {\usepackage{color}}{}
\makeatother
\usepackage{babel}

\usepackage{stmaryrd}
\usepackage{epsfig}
\usepackage{amsfonts}
\usepackage{amsmath}
\usepackage{graphicx}
\usepackage{color}
\usepackage{amssymb,amsmath,psfrag,slashed,graphicx}
\usepackage[normalem]{ulem}

\def \nn {\nonumber}

\allowdisplaybreaks

\begin{document}

%%%%%%%%%%%%%%%%%%%%%%%%%%%%%%

\title {{Universal modified function in conjunction with the short range correlation effect to extract the nuclear $xF_3^A$ structure function}}
\author{ A.~Mirjalili $^{1}$ \orcidA{}}
%\author{ A.~Mirjalili$^{1}$}
\email{a.mirjalili@yazd.ac.ir\;(corresponding author)}

\author{M.~Akbari Ahmadmahmoudi$^{1}$ \orcidB{}}
%\email{marzyehakbari088@gmail.com}

\author{H.~Abdolmaleki$^{2,3}$ \orcidC{}}
%\email{Abdolmaleki@ipm.ir}

\author{M.~M.~Yazdanpanah$^{4}$ \orcidD{}}
%\email{myazdan@uk.ac.ir}

%\newline\\

\affiliation {\vspace{.25 cm}
$^{(1)}$Physics Department, Yazd University, Yazd, Iran\\ \\
$^{(2)}$Department of Physics, Faculty of Science, Malayer University, Malayer, Iran\\ \\
$^{(3)}$School of Physics, Institute for Research in Fundamental Sciences (IPM),Tehran, Iran\\ \\
$^{(4)}$Physics Department, Shahid Bahonar University of Kerman, Kerman, Iran
}

\date{\today}

%
%%%%%%%%%%%%%%%%%%%%%%%%%%%%%%%%%%%%%%%%%%%%%%%%%%%%%%%%%%%%%%%%%%%%%%%%%%%%%%%%%%%%%%%%%%%%%%%%%%%%%%%
\begin{abstract}\label{abstract}
 Parton distribution functions (PDFs) are comprehensive and not reliant on the process. They are affected by nuclear matter during nuclear scattering process. As a recent approach, in order to study the nuclear PDFs (nPDFs), the nucleon pair PDFs are utilized to describe parton distributions in the nucleon pair which are confined to a nucleus. Nucleon pair PDFs  stem from nucleon-nucleon correlation  which is called short range correlation (SRC) and  are proportional to common nucleon PDFs. In this regard a modified universal  function is constructed  which provides  a test for SRC  in (neutrino)-nucleus scattering. In fact we can show that the modification of the structure function of  nucleons bound in atomic nuclei (known as  the EMC effect) are consistently accounted for within the frame work of a universal modification of nucleons in SRC pairs. In this article, based on the strategy which was introduced in Ref.\cite{nature}, we are investigating  to find the universality behaviour for the ratio of nonsinglet $xF_3^A$ nuclear structure function.
The numerical calculations performed within the CTEQ framework confirm the universality feature of the concerned ratio, as has been established for the ratio of the structure function $F^A_2$ in Ref. \cite{nature}. Following that we reformulate the $R_{EMC}$ nuclear weight function which relates the free and bound structure function in terms of modified universal function. It makes us a possibility to achieve $R_{EMC}$ and finally $xF_3^A$ nuclear structure function for each nucleus only by considering the specified feature of nucleus.
The results are compared with the nuclear $xF_3$ structure function which are computing from bound PDFs, based on some parameterizations models. The findings, considering the SRC effect, demonstrating qualitative agreement  with nCTEQ15 and EPPS21 parametrization models and also the available experimental data.
\end{abstract}

%\pacs{12.39.-x, 14.65.Bt, 12.38.-t, 12.38.Bx}
\maketitle

\section{Introduction}\label{Introduction}
The original idea of having nuclear effects in PDFs was driven by data in DIS measurements performed by the European Muon Collaboration (EMC) \cite{Aubert:1983xm}. The initial expectation was that physics at GeV scale would be insensitive to the nuclear binding effects which are typically on the order of several MeV scale. However, the collaboration discovered the per-nucleon deep inelastic structure function in iron is smaller than that of deuterium in the region $0.3<x<0.7$, here $x$ is the Bjorken variable. This phenomenon is known as EMC effect and has been observed for a wide range of nuclei \cite{Arneodo:1988aa,Seely:2009gt}. EMC effect indicates that quark PDFs in nucleon are modified, breaking down the scale separation between nucleon structure and nuclear structure. Although the understanding of how the quark-gluon structure of a nucleon is modified by the surrounding nucleons has been brought to a whole new level, one should note that there is still no consensus as to the underlying dynamics that drives this effect even after more than four decades since its discovering.

Currently, one of the leading approaches for describing the EMC effect is: nucleons bound in nuclei are unmodified, same as ``free'' nucleons most of the time, but are modified substantially when they fluctuate into short range correlation (SRC) pairs. The SRC describes the probability that two nucleons are close in coordinate space, as a result of nontrivial nucleon-nucleon interactions in nucleus. The connection between SRC and EMC effects has been extensively investigated in nuclear structure function measurements \cite{Hen:2014nza,Duer:2018sby}. { A linear relation between the magnitude of the EMC effect  and SRC scale factor, measured in electron DIS  at $0.3<x<0.7$ was proposed in \cite{Weinstein:2010rt}.} This striking linear relation suggests that { the EMC  and SRC effects} both stem from the same underlying physics such as high local density and high momentum nucleons in nuclei. This relation, if finally established, shall provide a unique method to study nuclear structure physics, see some of recent developments in Refs. \cite{Lynn:2019vwp,Hatta:2019ocp}. One of the key aspects of SRC is the universality, where the partonic structure from the correlated nucleon-nucleon pair is same for all kinds of nuclei, thus a universal modification function can be deduced. This function will be useful for testing QCD symmetry breaking mechanisms and for distinguishing nuclear physics effects from beyond the standard model effects in neutrino scattering experiments.

The robust linear correlation between the strength of the EMC and the SRC scale factor in nuclei indicates possible modifications of the quark PDFs occur in nucleons which are inside SRC pairs. In this connection, neutrino-nucleus DIS process is an ideal platform for testing nucleon structures and SRC interpretation of the EMC effect. It is sensitive to the quark distributions, especially for $u$ and $d$ quark (antiquark), they contain crucial information on nuclear effects which should be distinguished from beyond the standard model effects \cite{Grover:2018ggi,Li:2018wut,Yang:2020sos}.

{ This article  is organized as it follows. In Sec.{\ref{NNS}} we give a review on neutrino nucleus scattering and discuss how $xF_{3}^{A}$ nuclear structure function can be obtained in terms of free valence densities. In continuation in Sec.{\ref{Epps21}} the required description for the effect of SRC is given. Following that a modified weight function is obtained and some results are presented which indicate the universality property for different nuclei.  In Sec.{\ref{Epps}} a method is introduced in which the nuclear weight function for different nuclei is obtained in terms of universal weight function. Based on it the results for nuclear $xF_3^A$ structure function, including iron and lead nuclei, are presented which involve good agreement with the available experimental data  and  grid data points of some parametrization model such as CTEQ model. Last section, Sec.{\ref{Con}}, is allocated to our conclusion.}

\section{The requirements in neutrino nucleus DIS }\label{NNS}
\begin{figure}
\includegraphics[width=0.6\columnwidth]{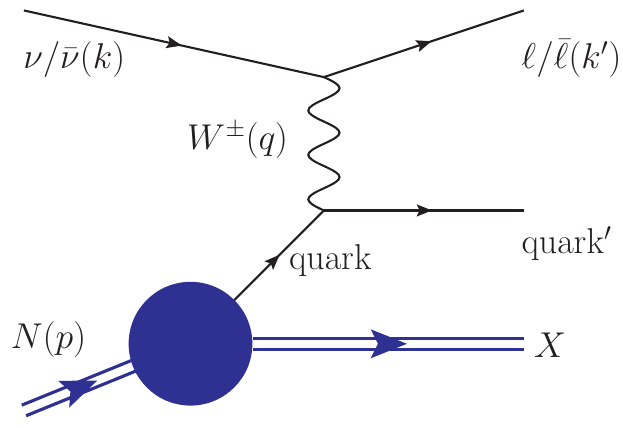}
\caption{A schematic diagram for neutrino charged-current DIS process, quoted from \cite{Dia}.}
\label{CC_DIS}
\end{figure}

In this section a review on neutrino nucleon DIS is given which lead us finally to { nucleus scattering process}. Neutrino charged-current DIS illustrates the process where a neutrino interacts with a quark in the nucleon through the exchange of a $W^{\pm}$ boson with momentum $q$, resulting in a corresponding lepton and hadron in the final state, which can be expressed as:
\begin{eqnarray}
  \nu_\ell(k)/\bar\nu_\ell(k)+N(p)\to \ell(k')/\bar\ell(k')+X(p') \,,
\end{eqnarray}
where the $k, p, k', p' $  { variables} indicate the four momentum of incoming and outgoing  particles.  A diagram, describing  neutrino charged-current DIS process is presented in Fig.\ref{CC_DIS}.

The neutrino (antineutrino) scattering cross sections from nucleus with mass number $A$ is given by \cite{Diemoz}:
\begin{eqnarray}\label{diff_for_nuandbarnu1}
  &&\left(\frac{d \sigma}{d x d y}\right)^{\nu {(\bar\nu)} A}=\frac{G_{F}^{2} M_{N} E_{\nu {(\bar\nu)}}}{\pi\left(1+Q^{2} M_{W}^{2}\right)^{2}}\nn\\
   &&\left[(1-y+\frac{y^{2}}{2})F_{2}^{\nu {(\bar\nu)} A}(x,Q^2)
  \pm  x y \left(1-\frac{y}{2}\right) F_{3}^{\nu {(\bar\nu)} A}(x,Q^2)\right] \;.\nn\\
\end{eqnarray}
Here $x\equiv Q^2/(2p\cdot q)$ is the Bjorken scaling variable and  $y\equiv(2p\cdot k)/(2p\cdot q)$ is the inelasticity variable where the  squared four momentum transfer is expressed by $Q^2\equiv -q^2$. The plus and minus signs in above are related respectively to neutrino and antineutrino scattering. The $M_n$ and $M_W$ are referred to nucleon and boson gauge mediate masses.

In present study we concentrate on nonsinglet  $xF_3^A$ { as the nuclear structure function that is an average function,} given by:
\begin{eqnarray}
xF_3^{A}(x,Q^2)=\frac{1}{2}\left(xF_3^{\nu A}(x,Q^2)+ xF_3^{\bar \nu A}(x,Q^2) \right)
\end{eqnarray}
{The choice to select this structure function is due to the significant cancellation of contributions from sea quarks, as discussed bellow, resulting in a focus mainly on valence quark effects. These considerations might also apply to $F_2$ structure functions in the limited large-x region where sea quarks and gluon contributions are negligible.}

{ The $xF_3^A$  nuclear structure functions can be written in terms of nuclear  parton densities as it follows \cite{Devenish}}
\begin{eqnarray}
xF_3^{\nu A} & =& 2x(d^A + s^A - \bar u^A - \bar c^A)~,   \nonumber   \\
xF_3^{\bar \nu A} & =& 2x(u^A + c^A - \bar d^A - \bar s^A)~.
\end{eqnarray}
{{ Within assumption $s^A=\bar s^A$ and $c^A=\bar c^A$ one will arrive at:}
\begin{eqnarray}
xF_3^{A}(x,Q^2)=xu_v^A(x,Q^2)  + xd_v^A(x,Q^2) ~.
\label{eq5}
\end{eqnarray}

Essentially, the nuclear parton density function (nPDFs) for a nucleus  can be represented as \cite{T-K-A}
\begin{equation}
xq_{v}^A(x,Q_0^2)=\frac{Z}{A}\;xq_{v}^{p/A}(x,Q_0^2) +\frac{A-Z}{A}\;xq_{v}^{n/A}(x,Q_0^2)\;,
\label{eq:pdfsnucleus}
\end{equation}
where $A$ and $Z$ refer respectively to the mass  and atomic number, while $p$ and $n$ denote proton and neutron inside the nuclei.
In the above equation, $xq_{v}^{p/A}$  and $xq_{v}^{n/A}$ represent valence PDFs of a bound
proton and neutron in the nucleus $A$.

The nuclear effects may stem from the modifications of PDFs due to nuclear interactions. For the nonsinglet QCD analysis, this adjustment establishes a link between the bound valence PDFs in the nucleus $A$ and the free valence PDFs in the proton as
\begin{equation}
xq_{v}^{p/A}(x,Q_0^2)= R^A_{q({v})}(x,A,Z)\;xq_v(x,Q_0^2)~,
\label{eq:nuclear}
\end{equation}
in which $q_v=u_v, d_v$. Here $R^A_{q({v})}(x,A,Z)$ represents the nuclear {{ weight} function that varies with the nucleus, while $xq_v(x,Q_0^2)$ denotes the valence PDFs in the free proton.

Various parametrization models are available for determining these nuclear weight functions. In this work, we utilize a novel approach based on short-range correlations (SRC) between nucleon pairs, which provides a physically motivated framework to quantify nuclear modifications in terms of correlated nucleon pairs. Next section is allocated to this subject.

\section{Short rang correlation and modified weight function}\label{Epps21}
The internal quark-gluon substructure of nucleons was originally expected to be independent of the nuclear medium because quark interactions occur at shorter-distance and higher-energy scales than nuclear interactions. However, DIS measurements indicate that quark momentum distributions in nucleons are modified when nucleons are bounded in atomic nuclei, breaking down the scale separation between nucleon structure and nuclear structure.  On this connection, as mentioned earlier, an amazing approach, based on short range correlations  effect is available to explain the influence of nuclear media on  nucleon structure.  SRC is in fact  brief fluctuations of nucleons in nuclei to form pairs with high relative momentum. The study of SRC is a broad subject, covering a large body of experimental and theoretical work, as well as phenomenological studies \cite{LBW,JA,Atti,Hen,JC,OH}.

\begin{figure}
\includegraphics[width=0.9\columnwidth]{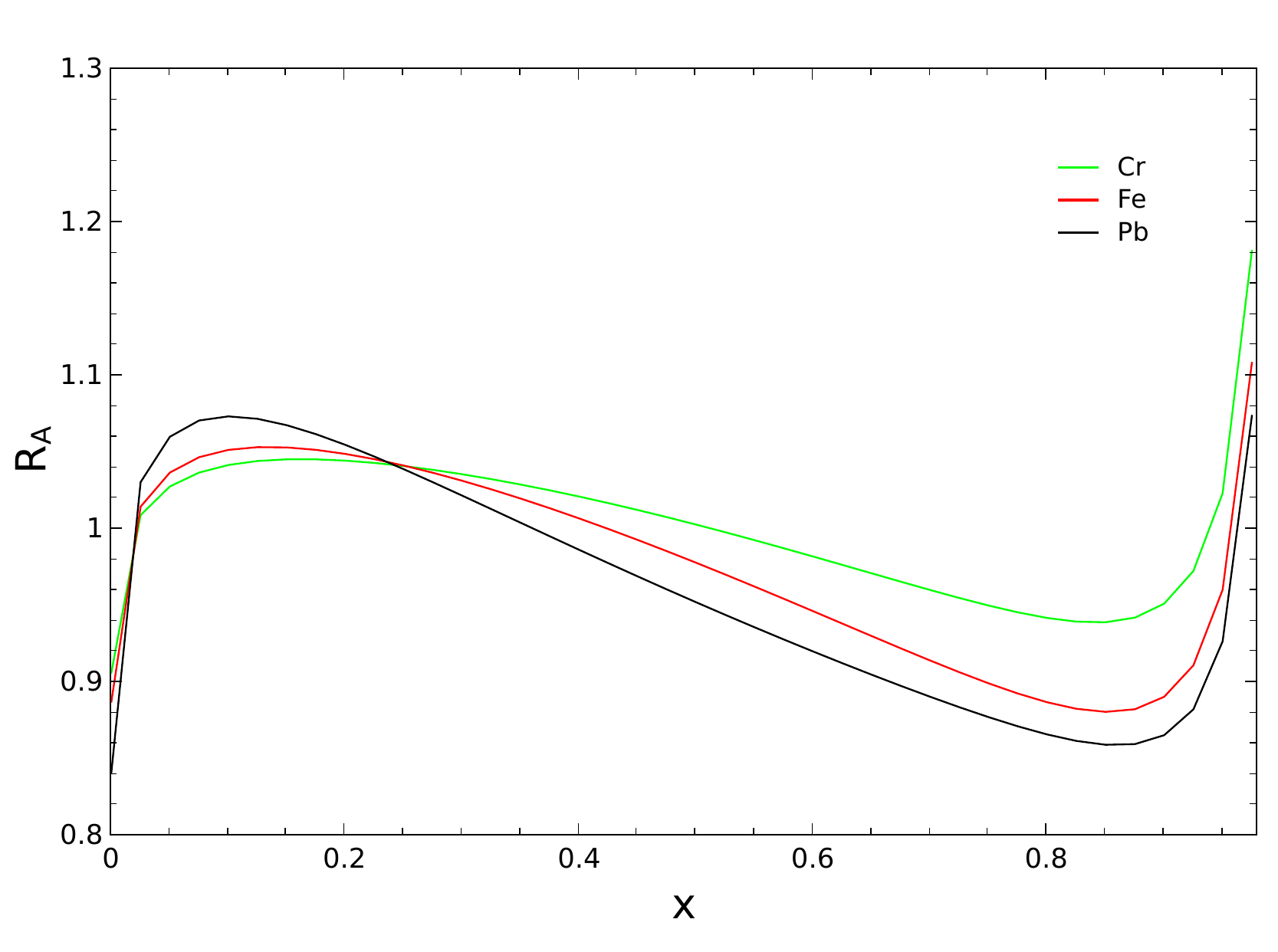}
\caption{{ Weight functions for different carbon, iron and lead nuclei, resulted from EPPS21 parametrization \cite{EPPS21}.}}
\label{Ratio-Dssz}
\end{figure}

\begin{figure}
\includegraphics[width=0.9\columnwidth]{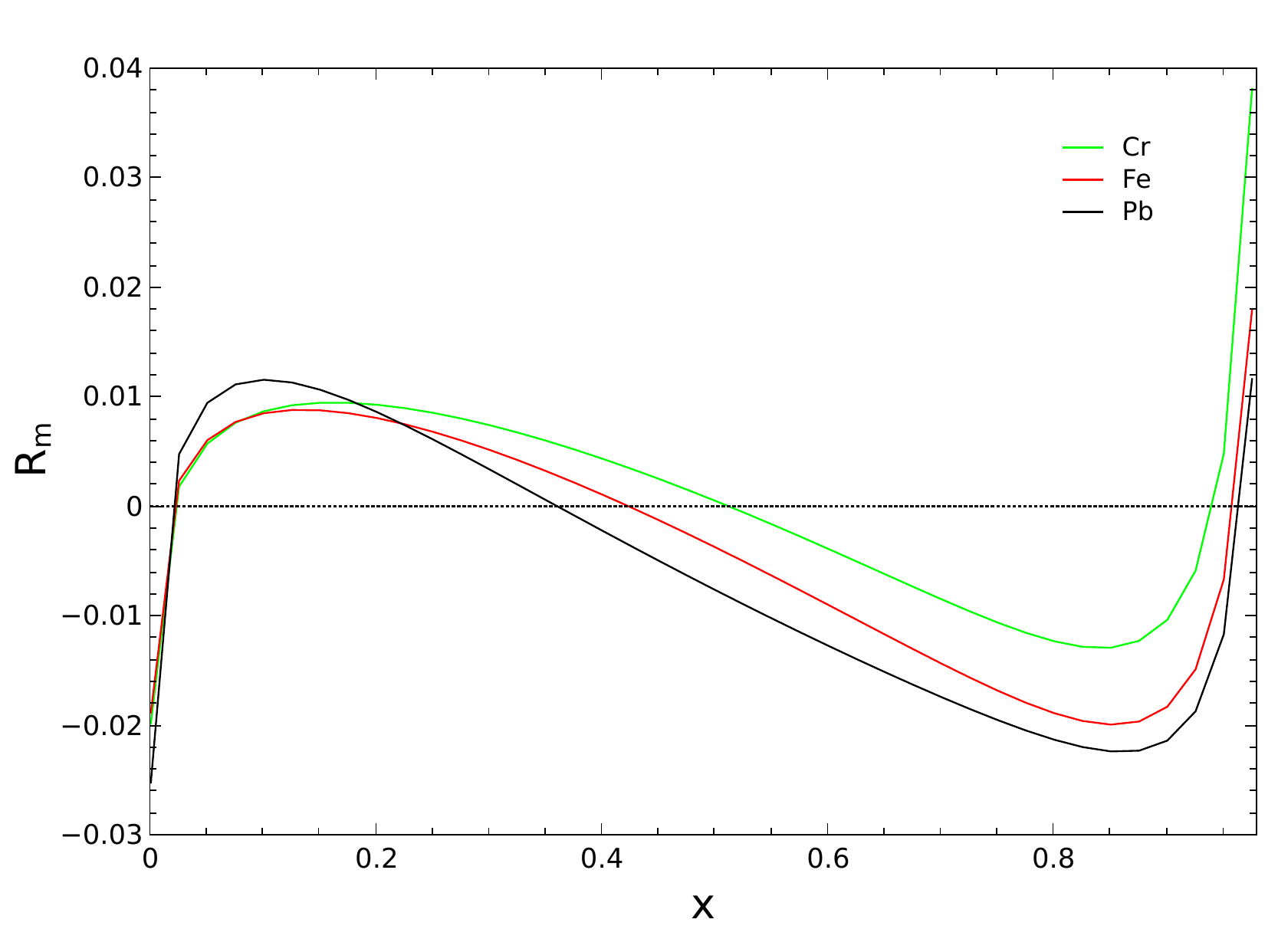}
\caption{{ Modified weight functions for different carbon. iron and lead nuclei, based on Eq.(\,\ref{Mod-Wei}).}}
\label{Ratio-Uniii}
\end{figure}

\begin{figure}
\includegraphics[width=0.9\columnwidth]{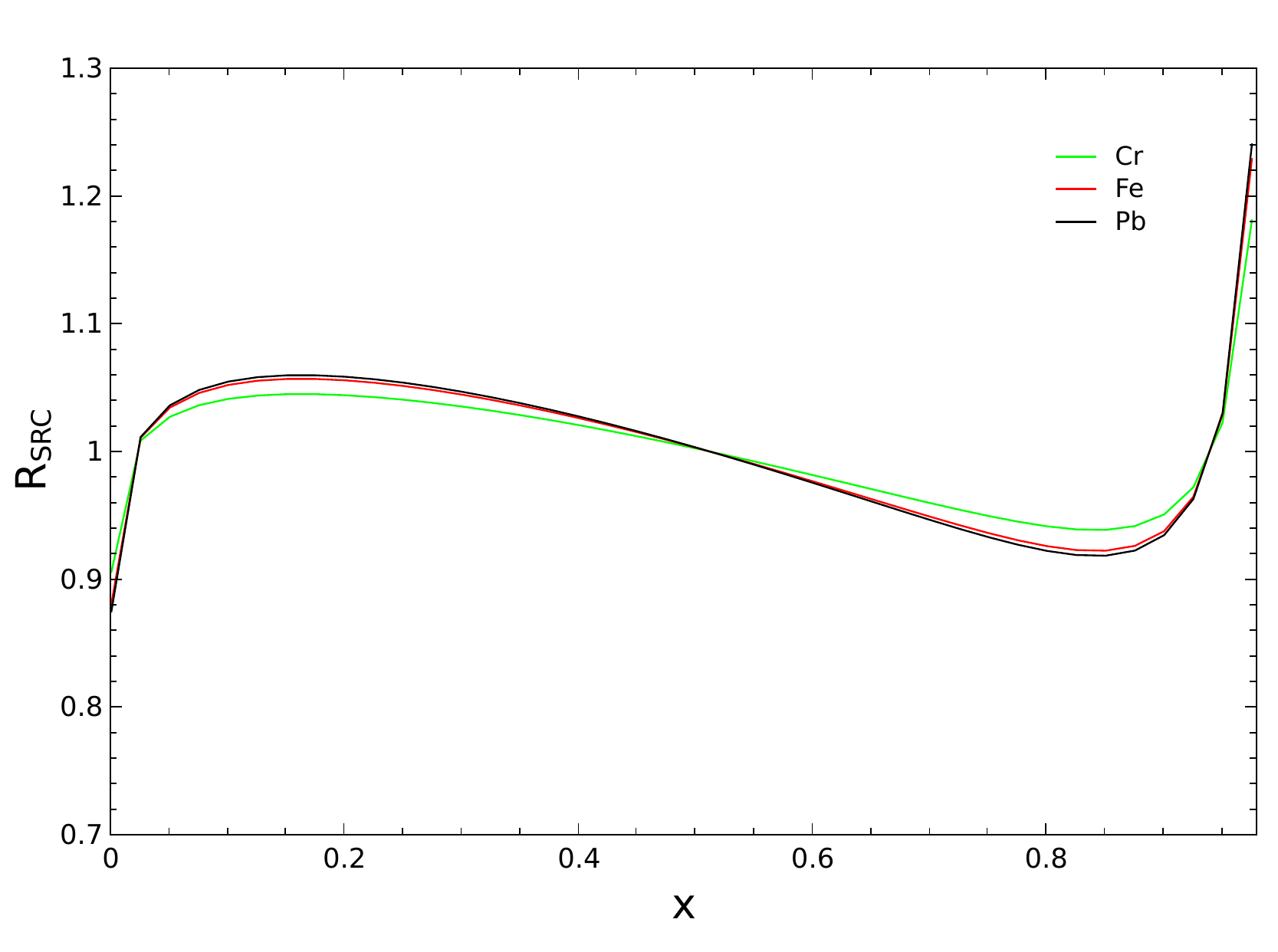}
\caption{{ Improved weight functions for different carbon, iron and lead nuclei, based on Eq.(\,\ref{use_of_uni_function}).}}
\label{Ratio-Src}
\end{figure}

A fundamental aspect of SRC is its universality, where the partonic structure of the correlated nucleon-nucleon pair is same across all types of nuclei, allowing for the derivation of a universal modification function. The neutrino-nucleus DIS process provides a solid basis for assessing the SRC evaluation of the EMC effect. Driven by the correlation between the EMC effect and the SRC scale factor, and in light of the SRC's universality, the distributions of $u$ and $d$ quarks in the EMC region can be modeled by presuming that all nuclear modifications stem from the nucleon-nucleon SRCs as outlined \cite{Seg,Hua}:
\begin{eqnarray}\label{para}
 f_{u_v(d_v)}^{p/A}\left(x, Q^{2}\right)=\frac{1}{Z}\left[Z f_{u_v(d_v)}^{p}\left(x, Q^{2}\right) \right. \nonumber \\
\left.+n_{s r c}^{A} \delta f_{u_v(d_v)}^{p}\left(x, Q^{2}\right)\right] \,.
\end{eqnarray}
Here $n^A_{src}$ represents number of neutron-proton pairs in nucleus $A$ and the subscript $v$ in $f_{u_v}$ and $f_{d_v}$ means distributions of valence quarks since experimental results pointed to the EMC effect being due to a change in the valence quark distributions. Finally $\delta f_{u_v(d_v)}^{p}$ represents the difference between $u(d)$ valence quark distribution in the SRC pair and in the free proton.

{It should be noted that the \(1/Z\) term in Eq.(\ref{para}) depends on the proton PDF being the relevant one, leading to the nuclear modification due to the SRC scaling factor. Owing to the prevalence of the p-n configuration in the SRC pairs, the Z factor inside the bracket of Eq.(\ref{para}) is also emphasized instead of the other factor such as \(A\). It is clear that if additional PDFs, besides the proton PDF, play a significant role in SRC pairs, then Eq.(\ref{para}) should  be modified accordingly with the factors and other terms included.}

Substituting Eq.(\ref{para}) into Eq.(\ref{eq:nuclear}), we will arrive at
\begin{eqnarray}\label{unii}
  \frac{\delta f_{u_v(d_v)}^{p}\left(x, Q^{2}\right)}{f_{u_v(d_v)}^{p}\left(x, Q^{2}\right)}=\frac{R_{u_v(d_v)}^{A}\left(x, Q^{2}\right)-1}{n_{src}^A/Z_A} \,.
\end{eqnarray}
The left hand side of above equation  is assumed to be nucleus-independent, meaning that the right-hand side should be a universal function which does not depend on the type of nucleus.

{Considering Eq.(\ref{unii}) for two different  $A$ and $B$ nuclei which includes SRC effect, one will arrive at the following relation for weight function of these two  nuclei \cite{Hua}:}
\begin{eqnarray}\label{use_of_uni_function}
  R_{u_v(d_v)}^{A}\left(x, Q^{2}\right)&=&\frac{n_{s r c}^{A} / Z_{A}}{n_{s r c}^{B} / Z_{B}}\left(R_{u_v(d_v)}^{B}\left(x, Q^{2}\right)-1\right)+1 \nn\\
  &=& \frac{a_2^A}{a_2^B}\frac{Z_{B}}{Z_{A}}\frac{A_A}{A_B}\left(R_{u_v(d_v)}^{B}\left(x, Q^{2}\right)-1\right)+1 \,. \nonumber\\
\end{eqnarray}
Here $a_{2}^{A}=\left(n_{s r c}^{A} / A\right) /\left(n_{s r c}^{d} / 2\right)$ is the SRC ratio of nucleus $A$ with respect to that of deuteron which can be measured through  DIS cross section which includes the nuclear structure functions \cite{Fom}.
The above relation provides a way to  parameterize the quark distribution in nucleus $A$ in terms of that in  nucleus $B$.
In fact, based on the EPPS21 parametrization of quark nPDFs \cite{EPPS21}{ for instance carbon nucleus} and using Eq.(\ref{use_of_uni_function}), one can get quark nPDFs for iron and lead { nuclei}.

Following the strategy which has been utilized in \cite{Yan}, Eq.(\ref{use_of_uni_function}) can be extended to weight functions  for structure functions of  $A$ and $B$ nuclei and consequently the following relation is arising out:
 \begin{eqnarray}
  \!\!\!\!\!\!\!\! \frac{2Z_A}{A_A} \; \! \frac{R(xF_3^{A} ;x,Q^2)-1}{a_2^{A}} \!&=&\! \frac{2Z_B}{A_B} \; \! \frac{R(xF_3^{B} ;x,Q^2)-1}{a_2^{B}}.  \label{Uni-Wei}
\end{eqnarray}}}

Hence the universality of SRC can be illustrated more specifically by introducing one kind of modified weight function as it follows \cite{Yan}:
\begin{eqnarray}
  \!\!\!\!\!\!\!\! R_{m}(xF_3^{A};x,Q^2) \!&=&\! \frac{2Z_A}{A_A} \; \! \frac{R(xF_3^{A} ;x,Q^2)-1}{a_2^{A}} \; . \label{Mod-Wei}
\end{eqnarray}
{In the process of deriving Eq.(\ref{Uni-Wei}) or Eq.(\ref{Mod-Wei}) from Eq.(\ref{use_of_uni_function}), it is important to observe that a factor of 2 appears. This feature would be confirmed by isoscalar nuclei where \(2Z/A\) is equal to 1, as expected.  In fact this feature is arising due to isospin symmetry for \(xF_3^{\nu p}\) and \(xF_3^{\nu n}\) from proton and neutron targets together with a similar status with an anti-neutrino beam directed at these targets. It is essential to note that the final structure function is obtained by averaging the separate structure functions.}

The modified weight function, resulted from Eq.(\ref{Mod-Wei}), is plotted in Fig.\,\ref{Ratio-Uniii} for three different carbon, iron and lead nuclei {where  the numerical values for the $a_2^{A}$ ratio utilized in our calculations are: $a_2^{Cr}=4.49$, $a_2^{Fe}=4.80$ and $a_2^{Pb}=4.80$. These values have been quoted from the measured CLAS data presented in Table I of Ref.\cite{nature}.} It can be seen that the ratios of different nuclei  draw nearer to one another compared with those in Fig.\,\ref{Ratio-Dssz} which are raised from EPPS21 parametrization. The findings reinforce the theoretical hypothesis presented in Eq.\,(\ref{para}) and suggest that in the valence quark dominated region $0.3\!<\!x\!<\!0.7$, the EMC effect is  mainly attributed to SRC pairs.{ The depiction in Fig.\,\ref{Ratio-Src} utilizes Eq.\,(\ref{use_of_uni_function}), where the carbon nucleus serves as the basis for considering the other two nuclei, affirming a significant test of SRC universality.} It offers a fresh insight into how the comparatively long-range nuclear force affects the short-distance parton structure within the nucleon.

Given the sufficient theoretical evidence for SRC and universal behaviors concerning the nuclear weight function, the next section outlines a formalism to employ this universality and finally  a way to extract the $xF_3$ nuclear structure function for various nuclei.

\section{Nuclear $xF_3^A$ structure function, using the modified universal weight function}\label{Epps}
\begin{figure}
\includegraphics[width=0.9\columnwidth]{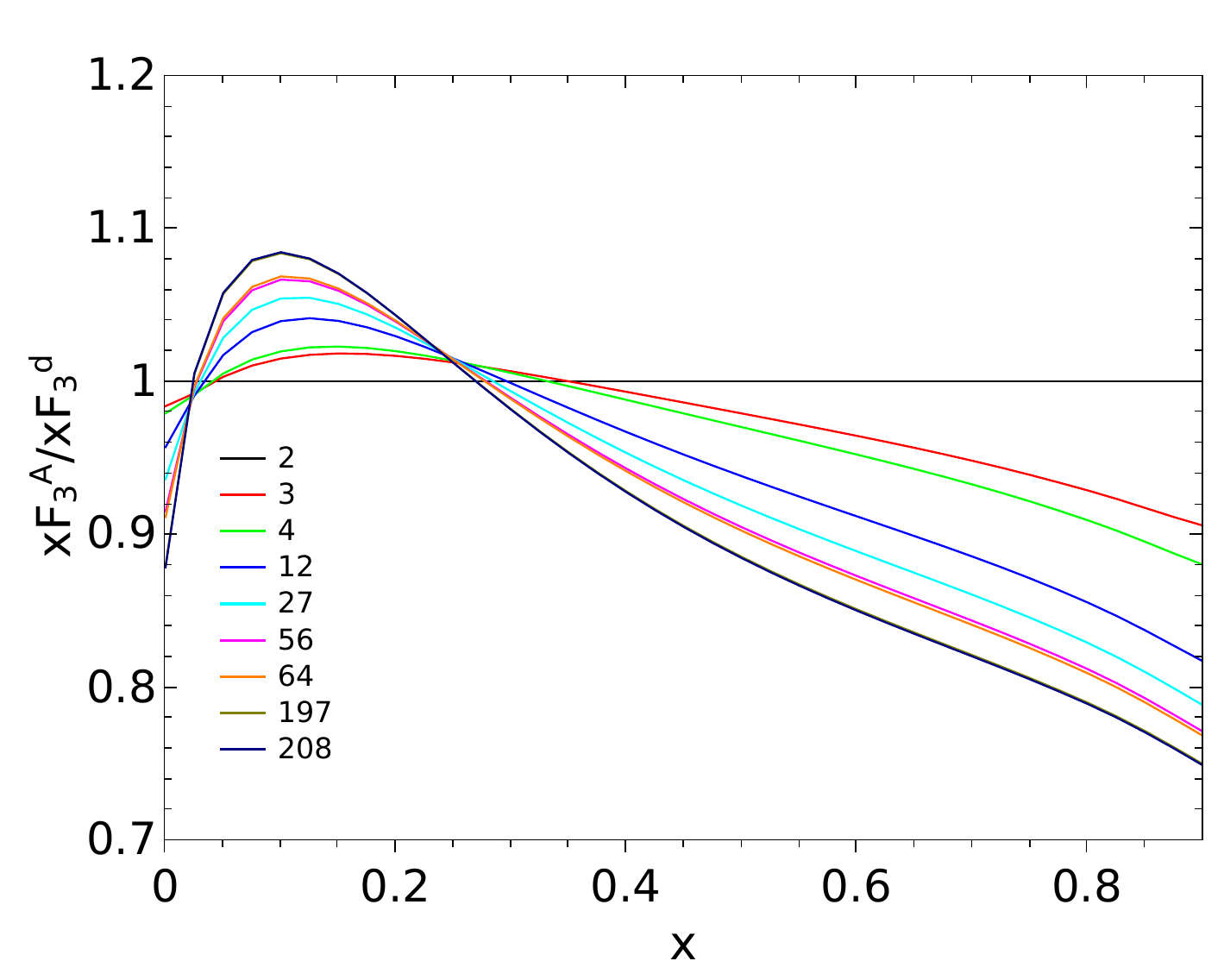}
\caption{The ratio of $xF^A_3/xF^d_3$  with respect to $x$-Bjorken scale for different nuclei at {$Q^2=10$ GeV$^2$.}}
\label{Ratio-EMC}
\end{figure}

Using the nuclear parton distribution functions (nPDF) from the nCTEQ15 \cite{Kovarik:2015cma} model and the ManeParse framework \cite{Clark:2016jgm}, a mathematica reader for PDFs, we compute the ratio of $xF^A_3/xF^d_3$ with respect  to the $x$-Bjorken scale for different nuclei, including Li, He, F, C, Al, Fe, Cu, Pb, and U at an energy scale of {$Q^2=10$ GeV$^2$}. The outcome has been depicted in Fig.\ref{Ratio-EMC}. It can be observed that there is some variability in the plotted ratio for various nuclei.\\

\begin{figure}
\includegraphics[width=0.9\columnwidth]{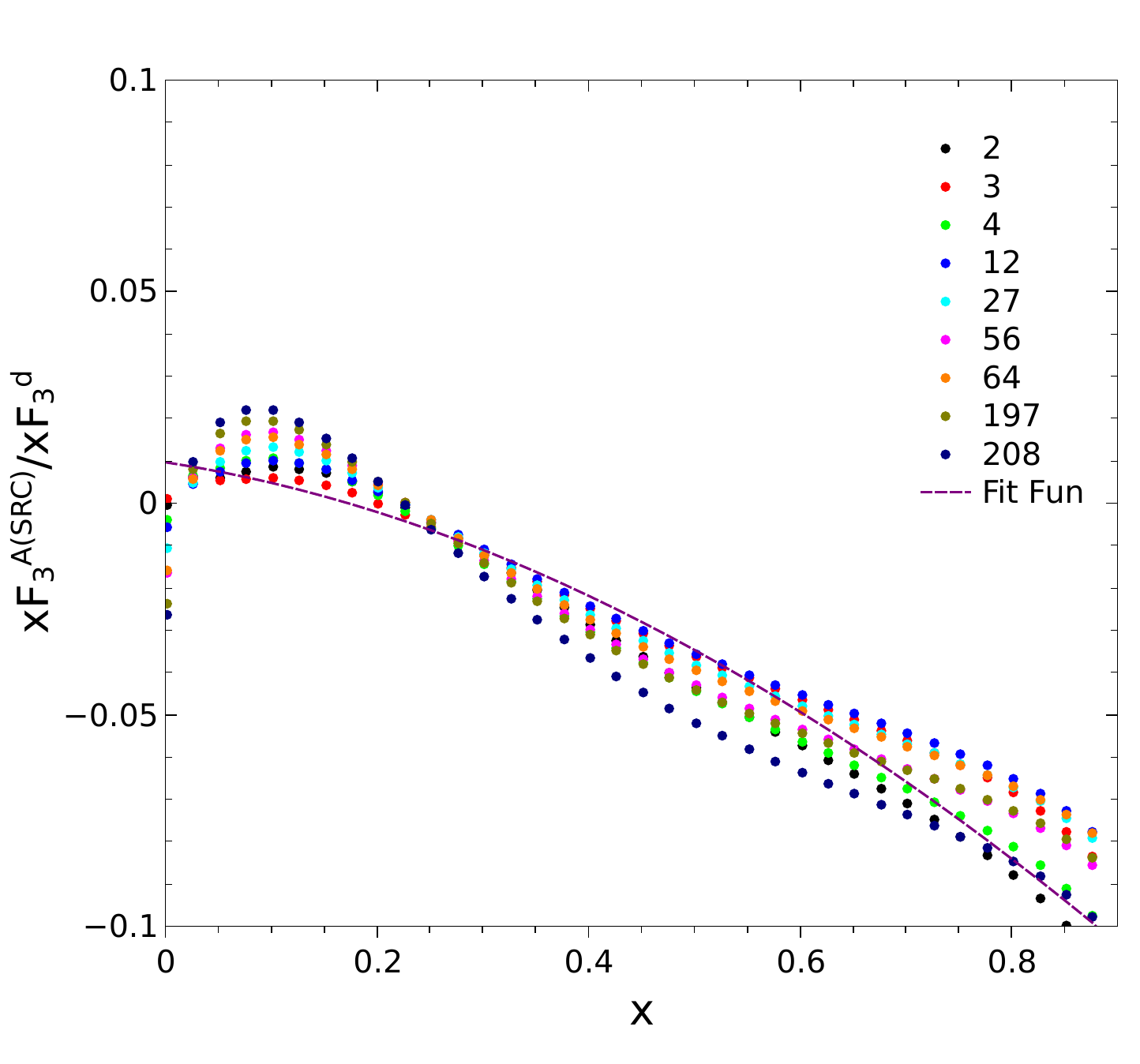}
\caption{The ratio of modification of SRC pair for $xF^A_3/xF^d_3$  with respect to $x$-Bjorken scale for different nuclei  at {$Q^2=10$ GeV$^2$.}}
\label{Ratio-Uni}
\end{figure}

Subsequently, using Eq.(2) from \cite{nature} but applied to the nuclear $xF_3$ structure function, we derive the ensuing relation:
\begin{eqnarray}
\frac{n^d_{SRC}(\Delta\; xF^p_3+\Delta\; xF^n_3)}{xF_3^d}=\frac{\frac{xF_3^A}{xF_3^d}-(Z-N)\frac{xF_3^p}{xF_3^d}-N}{\frac{A\;a_2}{2}-N}\;. \label{RatioN-EMCN}\nonumber\\
\end{eqnarray}
Since the left side of this equation, resulting from the SRC effect, is independent of any nucleus, we would expect that the right side possesses the same characteristic and demonstrates the necessary universality behavior. {Henceforth, we refer to the right side of Eq.(\ref{RatioN-EMCN}) as the modified universal function.} In Fig.\ref{Ratio-Uni}, we display the right side ratio of Eq.(\ref{RatioN-EMCN}) for various nuclei. As anticipated, the plots for various nuclei are converging and verify the expected universal behavior. By merging all data points  with respect to $x$-Bjorken variable for various nuclei and performing a fit on them, we obtain a curved line represented by black color in Fig.\ref{Ratio-Uni}
The functional shape of this curve, serving as a depiction of universal behavior for various nuclei, is utilized in other parts of our calculations.

\begin{figure}
\includegraphics[width=0.9\columnwidth]{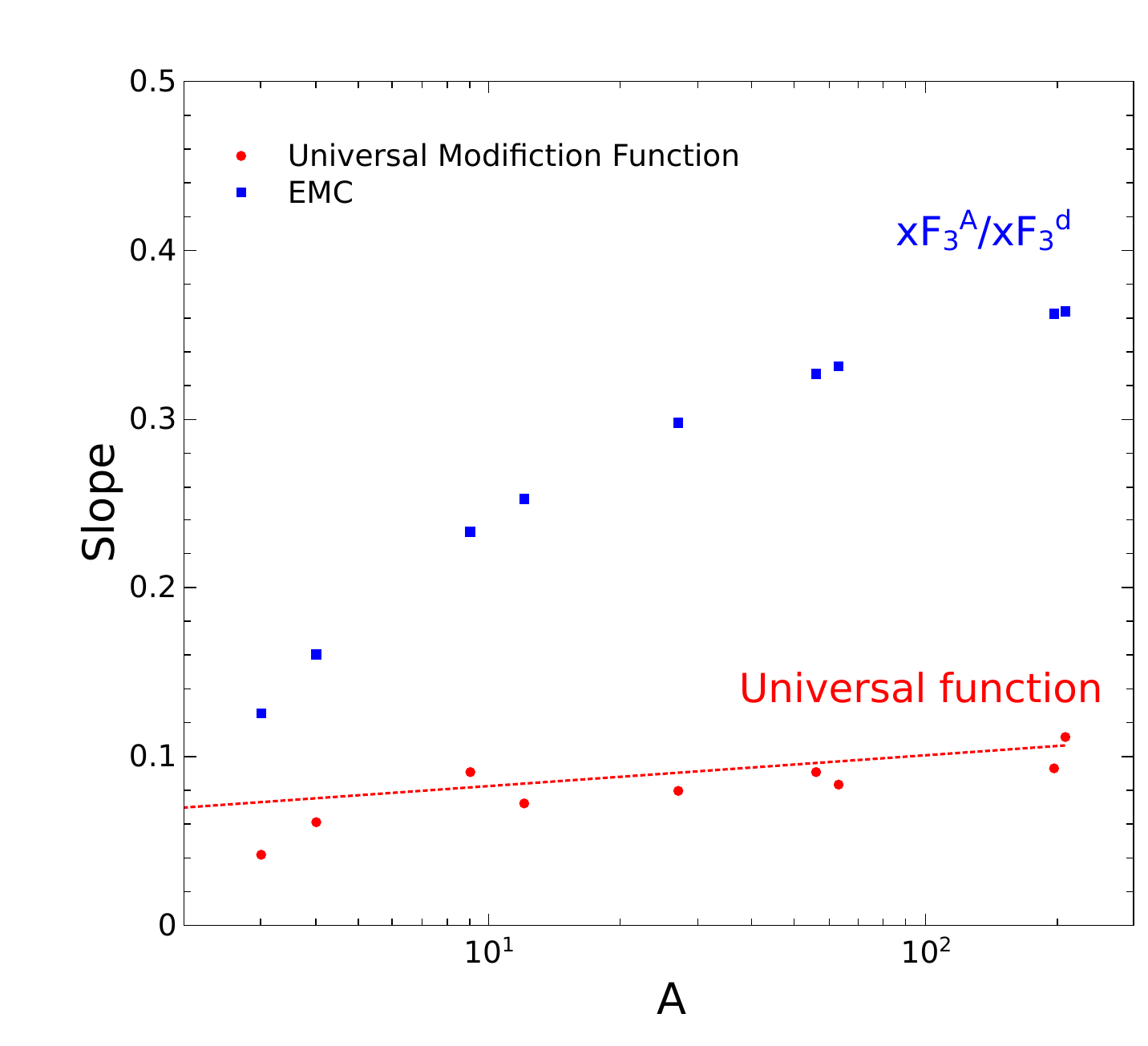}
\caption{EMC and universal modified slopes for $xF^A_3$  with respect to mass number $A$.}
\label{EMCslop-Unislop}
\end{figure}

One can also plot the slope of the EMC effect for the $xF^A_3/xF^d_3$ ratio in relation to the mass number $A$. We actually apply a linear fit to the $xF^A_3/xF^d_3$ ratio for specific nuclei concerning the $x$ Bjorken variable and derive its numerical slope in relation to the mass number $A$. Numerical values of EMC slope for various nuclei have been illustrated in Fig.{\ref{EMCslop-Unislop} by blue color. Numerical slopes for the modified universal function, using Eq.(\ref{RatioN-EMCN}), have been plotted in the same figure.{ The linear function related to mass number A that serves as the slope of the modified universal function is expressed in this way:
\begin{eqnarray}
S_f^u(A) = a_u+ b_u\;A
\label{Sfit}
\end{eqnarray}
where the numeric values for the fitted parameters are $a_u$=0.0695199\; and $b_u$=0.000178467\;. This function is depicted in Fig.{\ref{EMCslop-Unislop}} with a red dashed line.}}

The numerical values of modified slope with respect EMC slope represent more and less a constant value relative to vertical axes, confirming the existence of a universality characteristic for the modified slope. {This feature is also validated by the small numerical value of the $b_u$ parameter in Eq.(\ref{Sfit}).}

\begin{figure}
\includegraphics[width=0.9\columnwidth]{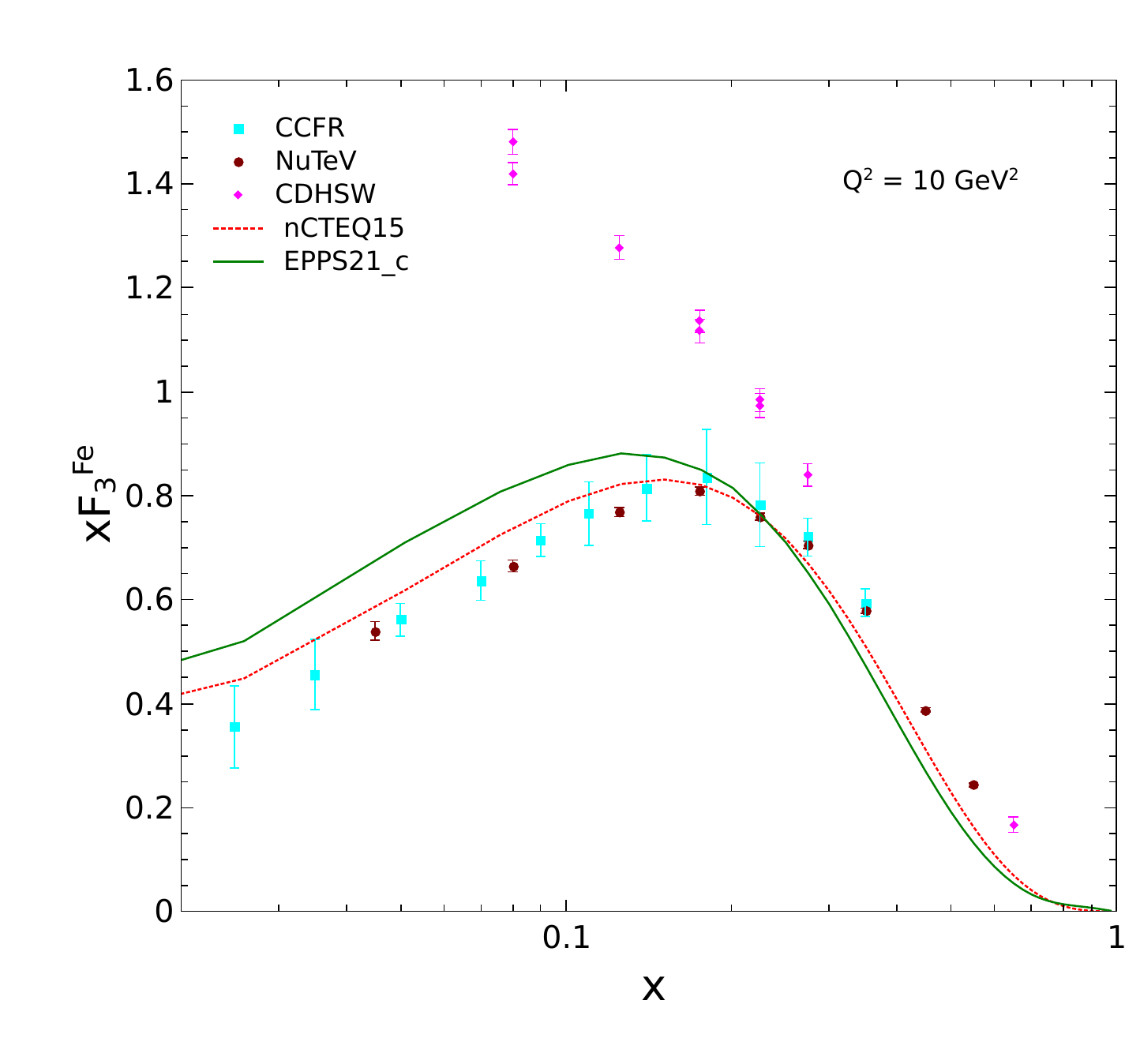}
\caption{{ Nuclear $xF^{Fe}_3$ structure function for iron nucleus, resulted from nCTEQ15 \cite{Kovarik:2015cma} and EPPS21 \cite{EPPS21}  parametrization model, presented by  dashed red curve and solid green  respectively. Comparison with the available experimental data \cite{CDHSW,CCFR,NuTeV} has also been done.}}
\label{Fig3-NNN}
\end{figure}

%These computations has also been done for different nuclear $xF_3$ structure functions. The results has been plotted in three Figs.(?,?,?).

{To more clearly demonstrate the SRC effect on modifying the nuclear structure function, we initially examine the $xF_3^A$ structure function for the iron nucleus, applying nCTEQ15 and EPSS21 parametrization models while excluding the SRC effect.  We actually use the following relation to calculate the $xF_3^A$ structure function
\begin{eqnarray}
xF_3^A=R_{u}^v\;xu_v+R_{d}^v\;xd_v \label{F3Ap}\;.
\end{eqnarray}

In this equation, the $R_{u}^v$ and $R_{d}^v$ weight functions for PDFs originate from the EPPS21 parametrization for the nuclear parton distribution function \cite{EPPS21}. The necessary $xu_v$ and $xd_v$ PDFs are derived from the nCTEQ PDFs parametrization \cite{Kovarik:2015cma}. By replacing the necessary functions on the right side of Eq.(\ref{F3Ap}), we can calculate $xF_3^{Fe}$. On the other hand, examining the left side of Eq.(\ref{F3Ap}), the $xF_3^{Fe}$ can be calculated directly, employing the nCTEQ nuclear PDFs. The outcomes of the left and right sides of this equation, which exclude any SRC effect, are represented by dashed and solid curves, respectively, in Fig.\ref{Fig3-NNN}.
{The data sets used  in this figure and other parts of this study derive from charged-current deep inelastic neutrino-nucleus scattering experiments, which provide sensitive probes of nucleon structure modifications in the nuclear environment. These measurements were performed using high-energy neutrino beams on heavy nuclear targets such as iron and lead. The structure function $xF_3(x,Q^2)$, primarily sensitive to valence quark distributions, was extracted over a broad kinematic range characterized by the Bjorken scaling variable $x$ and the four-momentum transfer squared $Q^2$.

This analysis focuses on the nuclear medium effects manifested as modifications to the structure functions, specifically targeting the EMC effect and nucleon-nucleon SRC. To ensure validity within perturbative QCD, only data satisfying the deep-inelastic region constraints  with $Q^{2} \geq 5\,\mathrm{GeV}^2$ and kinematic cut $W^{2} \geq 12.5\,\mathrm{GeV}^2$ were included. Unlike global QCD fits, this work does not consider extractions of fundamental parameters such as the strong coupling constant or higher twist effects but emphasizes phenomenological implications of the observed nuclear modifications.

The data sets employed include:

\textbf{CCFR \cite{CCFR}}: covering $x = 0.0075$ to $0.75$ and $Q^{2} = 1.3$ to $125.9\,\mathrm{GeV}^2$ with 116 initial points, reduced to 67 after kinematic cuts.

\textbf{NuTeV \cite{NuTeV}}: spanning $x = 0.015$ to $0.75$ and $Q^{2} = 1.26$ to $50.12\,\mathrm{GeV}^2$, with 75 original points and 59 passing cuts.

\textbf{CHORUS \cite{CHORUS}}: including 67 points over $x = 0.02$ to $0.65$ and $Q^{2} = 0.325$ to $81.55\,\mathrm{GeV}^2$, with 41 points retained for analysis.

\textbf{CDHSW \cite{CDHSW}}: providing 143 points within $x = 0.015$ to $0.65$ and $Q^{2} = 0.19$ to $196.3\,\mathrm{GeV}^2$, with 96 used after cuts.

These data provide a comprehensive and complementary coverage of the relevant kinematic region for investigating nuclear effects within neutrino deep inelastic scattering.

}}

{In this context, it is important to examine the relationship between Eq.(\ref{F3Ap}) and Eq.(\ref{eq5}). In fact Eq.(\ref{eq5}) expresses the general definition of the nuclear structure function using bound valence PDFs while Eq.(\ref{F3Ap}) explicitly realizes this definition by replacing those bound valence PDFs with their weighted forms based on EPPS21 nuclear modification factors and nCTEQ free-proton PDFs.}

In the next step, to realize practical results from the SRC behavior of nucleon pairs, we express the right side of Eq.(\ref{RatioN-EMCN}) for $xF^A_3$ using $R_f^u$ as follows:
\begin{eqnarray}
R_f^u=\frac{\frac{xF_3^A}{xF_3^d}-(Z-N)\frac{xF_3^p}{xF_3^d}-N}{\frac{A a_2}{2}-N}\;. \label{rfu}
\end{eqnarray}

{For computational purposes the selected fitting shape of this function which we called it  universal modified function, is as follows:
\begin{eqnarray}
R_f^u(x) = a + b\;x + c\;e^{d\;(1 - x)}\;,
\label{Ffit}
\end{eqnarray}
in which the acquired numerical values of the fitted parameters are
$
a=3.60462, \;b=-0.908723, \;c=-2.82168, \;d=0.242195.
$ As mentioned earlier this fitted function is corresponding to the  curved line in Fig.{\ref{Ratio-Uni}}.}\\

If in Eq.(\ref{rfu}) we make use of the following expression for $xF^A_3/F^d_3$
\begin{eqnarray}
\frac{xF_3^A}{xF_3^d}=\frac{xF_3^A}{xF_3^p}\frac{xF_3^p}{xF_3^d}=R_{EMC}\frac{xF_3^p}{xF_3^d}\;,\label{rfu-emc}
\end{eqnarray}
we can derive the EMC weight function, $R_{EMC}$, expressed in terms of $R_f^u$ as a universal modified function that incorporates the SRC effect based on the relation that follows:
\begin{eqnarray}
R_{EMC}=\frac{xF_3^A}{xF_3^p}=(Z-N)+\frac{N xF_3^d}{xF_3^p}+({\frac{A}{2}a_2-N})\frac{xF_3^d}{xF_3^p}R_f^u\;. \nonumber\\ \label{remc}
\end{eqnarray}

By applying the nuclear weight function from the equation above, we can derive the nuclear $xF_3^A$ structure function, as shown:
\begin{eqnarray}
xF_3^A=R_{EMC}\; xF_3^p \label{F3A} \label{xf3A}
\end{eqnarray}
In this equation, $R_{EMC}$ is derived from Eq.(\ref{remc}) and the structure function $xF_3^p$ is obtained from the nCTEQ15 parton distribution function (nCTEQ15 PDFs) \cite{Kovarik:2015cma}. The modified universal function, $R_f^u$, in Eq.(\ref{remc}) is substituted by  the functional form of fitted curve in Fig.\ref{Ratio-Uni}. All other parameters in this equation, including the SRC scale factor $a_2$, atomic $Z$, and mass number $A$, are provided, where the neutron number is determined by $N=A-Z$.

\begin{figure}
\includegraphics[width=0.9\columnwidth]{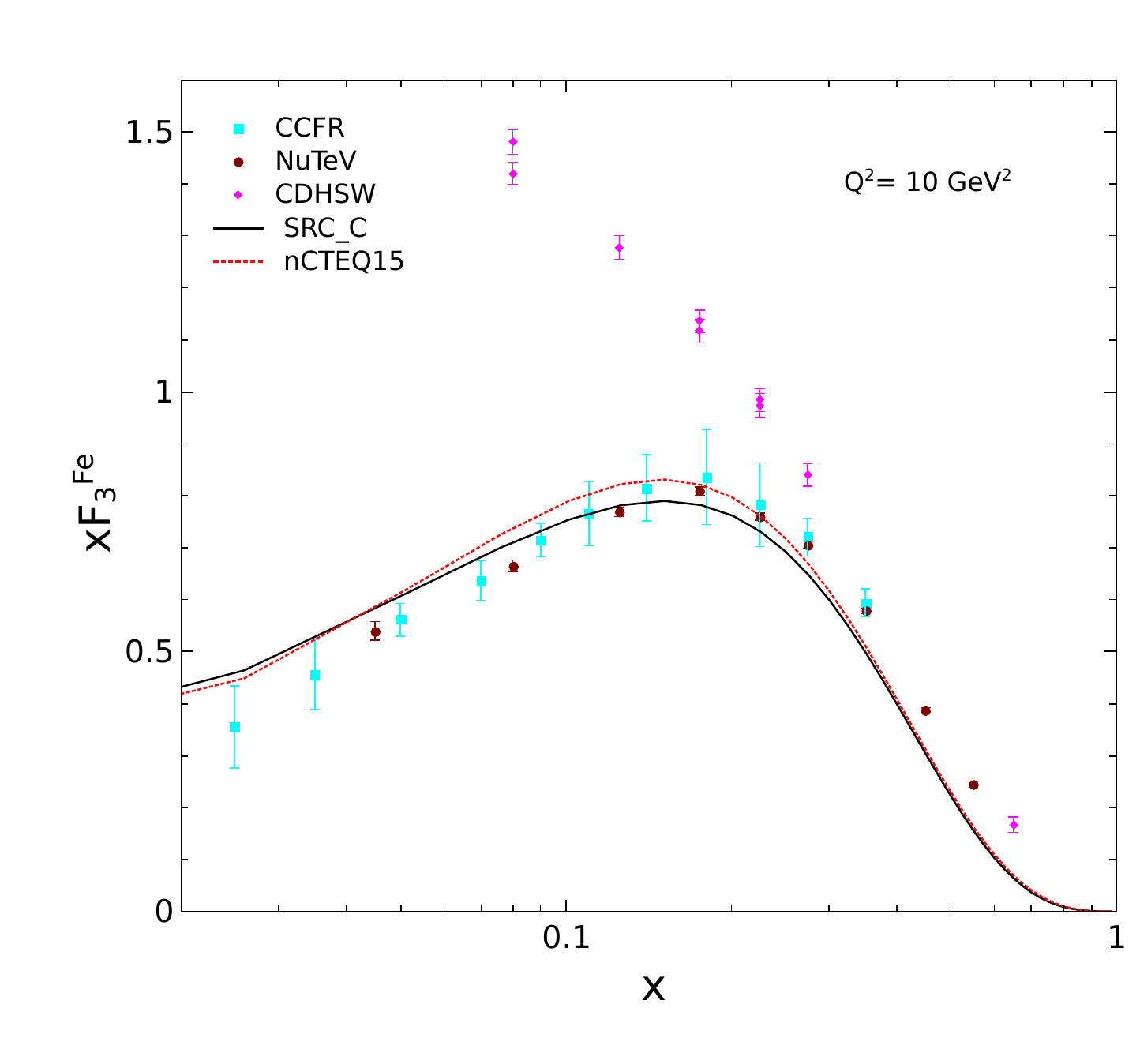}
\caption{Nuclear $xF^{Fe}_3$ structure function for iron nucleus, resulted from SRC modification (solid black curve) in comparison with nCTEQ15 parton distribution function \cite{Kovarik:2015cma} for  $xF^{Fe}_3$ (dashed red curve). Comparison with the available experimental data \cite{CDHSW,CCFR,NuTeV} has also been done.}
\label{Fig3-N}
\end{figure}

However, $xF_3^A$ in Eq.(\ref{F3A}) is calculated using the $R_{EMC}$ weight function that encompasses the SRC effect via the employed $R_f^u$ function, but it can be directly extracted from nCTEQ15 nuclear PDFs. The outcome for the iron nucleus is shown in Fig.\ref{Fig3-N}, demonstrating good agreement between the results of these two methods.

\begin{figure}
\includegraphics[width=0.9\columnwidth]{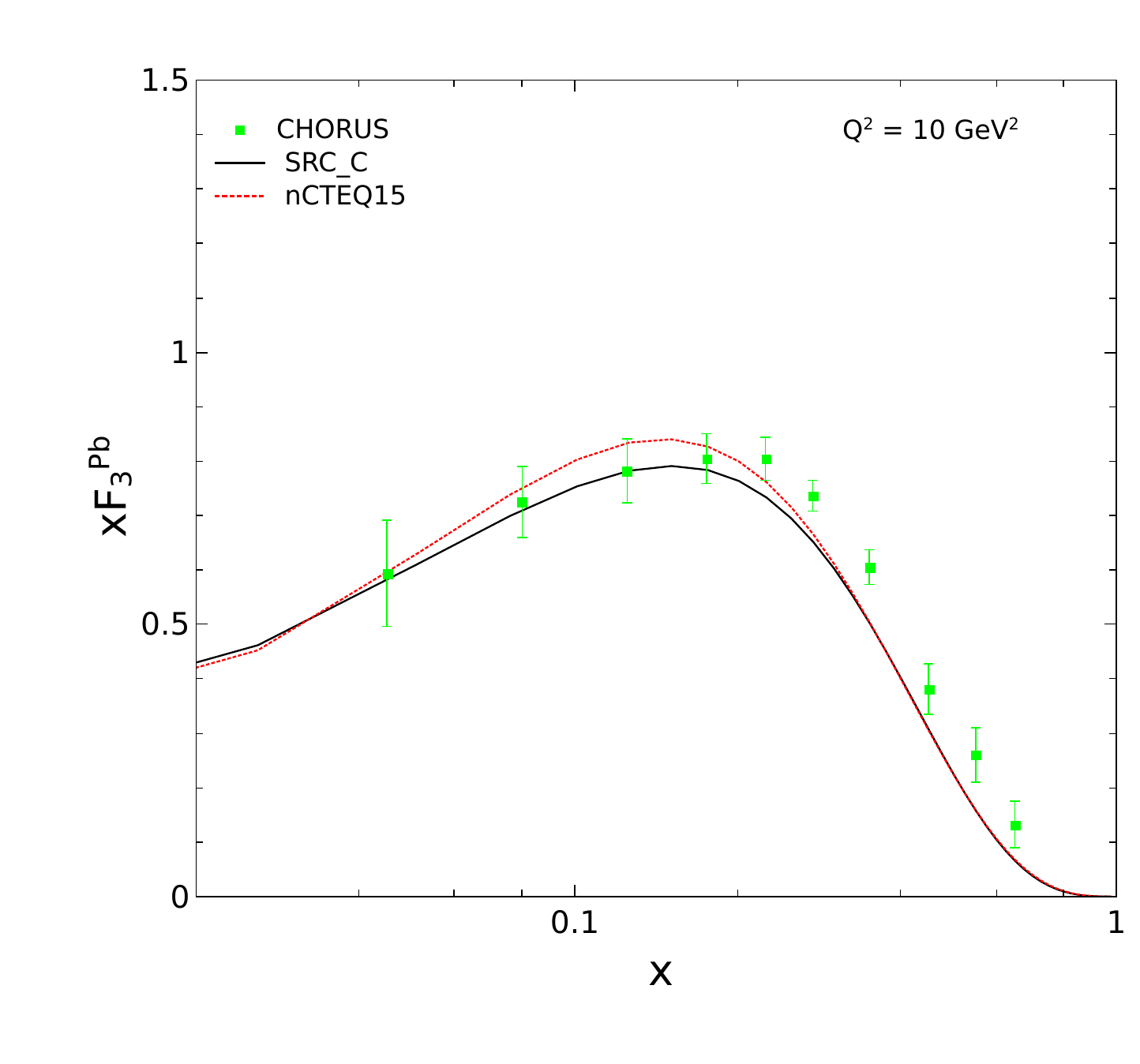}
\caption{Nuclear $xF^A_3$ structure function for lead nucleus, resulted from SRC modification (solid black curve), in comparison with $xF^A_3$ resulted form nCTEQ15 PDFs \cite{Kovarik:2015cma} (dashed red curve). Comparison with the available experimental data \cite{CHORUS} has also been done.}
\label{Fig3}
\end{figure}

{ The comparison of Fig.\ref{Fig3-NNN} and Fig.\ref{Fig3-N} clearly demonstrates that the SRC effect improves the outcome for the $xF_3^{Fe}$ structure function. The results of the SRC effect, depicted in Fig.\ref{Fig3-N} by the solid black curve, is in better agreement  with the avilable experimental data rather than the dashed red curve in both figures, derived from the nCTEQ15 model. It also shows better agreement with respect to the EPPS21 parametrization,  illustrated by the solid green curve in Fig.\ref{Fig3-NNN}.

{{{As can be seen in Fig.\ref{Fig3-NNN} and Fig.\ref{Fig3-N} the structure function $xF_3$ extracted from the CDHSW experiment \cite{CDHSW} does not fully agree with the results from the  CCFR \cite{CCFR} and NuTeV \cite{NuTeV} collaborations, especially at intermediate and low values of Bjorken-x. The CDHSW data tend to be higher in these x regions compared to the more recent NuTeV and CCFR measurements, which benefit from improved statistics, systematic corrections, and nuclear effects treatment. These differences are largely attributed to advancements in experimental techniques and nuclear corrections over time. Currently, NuTeV and CCFR results are considered more reliable for QCD analyses, while CDHSW data retain historical significance.}}

The accomplishment we attained is that we derived a weight function for various nuclei, shown in Eq.(\ref{remc}), which is easily applicable since modified universal function, $R_f^u$, can be used for every nucleus. Other variables in this equation are likewise readily obtainable for each nucleus. Based on this foundation, we utilize Eq.(\ref{F3A}) to extract $xF^A_3$ for the lead nucleus and present the findings in Fig.\ref{Fig3}, which is compared with available data and the nCTEQ15 outcome \cite{Kovarik:2015cma}. {{In fact, as for iron nucleus, the nuclear structure function $xF_3^A$ for the lead nucleus is derived, on one side, by applying $R_{EMC}$ in Eq.(\ref{remc}) alongside $xF_3^p$ from nCTEQ15 PDFs, and on the other side, by directly obtaining $xF_3^A$ for the lead nucleus from nCTEQ15 nuclear PDFs.}}

{In summary, using the precise form in Eq.(\ref{xf3A}), we could calculate the nuclear structure function in two equivalent ways:
\begin{enumerate}
    \item by inserting the \(R_{\mathrm{EMC}}\) function from Eq.(\ref{remc}) into the right-hand side of Eq.(\ref{xf3A}), employing nCTEQ15 PDFs for \(xF_3^p\), and
    \item by directly evaluating \(xF_3^A\) from the nCTEQ15 nuclear PDFs.
\end{enumerate}
Both approaches yield consistent results, as presented in Fig.\ref{Fig3-N} and Fig.\ref{Fig3} for Fe and Pb nuclei. The consistency between these two independent calculations confirms that the \(R_{\mathrm{EMC}}\) formulated in Eq.(\ref{remc}), together with the universal function \(R_f^u\), reliably reproduces nuclear modifications consistent with EPPS21 and nCTEQ15 frameworks.}

%The same procedure can be done for lead structure function which has depicted in  Fig.\ref{lead}. data is coming from  CHORUS  [??] group.
%\begin{figure}
%\includegraphics[width=0.9\columnwidth]{xF3-Fe.pdf}
%\caption{Nuclear $xF^A_3$ structure function for lead, resulted from SRC modification (blue color) in comparison with EPPS21 parametrization model (green color).}
%\label{lead}
%\end{figure}

\section{Conclusion}\label{Con}

Parton distribution functions play a crucial role in characterizing nuclear structures. Nonetheless, nucleon PDFs are unavoidably influenced by nuclear matter in the lepton-nucleus DIS process. Several experiments have additionally validated the fact that PDFs assessed in free nucleons and in the nuclei containing bound nucleons are remarkably different.

Several approaches exist to explore the nuclear PDF. One of them that is of note near to a decade, utilizes  the short range correlation (SRC) effect, which arises from the fluctuation of nucleon pairs. This phenomenon results in a characteristic that is  more and less universal across all nuclei, and we intended to employ it to extract the $xF_3$ nuclear structure function. To achieve this, following a short overview of nuclear DIS processes, we utilized the SRC effect to determine the modified weight function, as described by Eq.(\ref{Mod-Wei}) for carbon, iron, and lead nuclei. The outcome has been illustrated in Fig.\ref{Ratio-Uniii}. Subsequently, we applied Eq.(\ref{use_of_uni_function}) to derive the improved weight function for the three specified nuclei, depicted in Fig.\ref{Ratio-Src}, which notably demonstrates the universality characteristic derived from the SRC effect

In continuation, we introduced a technique for using the universality property of the weight function to derive the nuclear weight function for iron and lead, which can also be applied to other nuclei. To achieve this, we initially presented in Fig.\ref{Ratio-EMC} the ratio of $xF^A_3/xF^d_3$ structure functions as a function of the $x$-Bjorken scale for various nuclei at $Q^2=10$ GeV$^2$, derived from  nCTEQ15 nuclear PDFs. Subsequently, we illustrated in Fig.\ref{Ratio-Uni} the right-hand side ratio of Eq.(\ref{RatioN-EMCN}), which has been obtained from the modification of SRC pairs, and we anticipate it to demonstrate the necessary universality feature. The findings in Fig.\ref{Ratio-Uni} support this characteristic.

To further illustrate the universal behavior of various nuclei, based on the SRC characteristics of nucleon pairs, we displayed in Fig.\ref{EMCslop-Unislop} the slopes for the $xF^A_3/xF^d_3$ ratio, referred to as the EMC slope. In this figure, we similarly illustrated the slope of the right-hand side in Eq.(\ref{RatioN-EMCN}) as a universal modification slope. Since the ratios from SRC pairs in Fig.\ref{Ratio-Uni} for various nuclei are closing to each other, their slopes concerning $x$-values would be relatively constant. This reality is illustrated in Fig.\ref{EMCslop-Unislop}, showing universal modification slopes that indeed affirm the anticipated universal behavior across various nuclei.

{ Prior to examine the SRC effect, we presented the results for the iron $xF^A_3$ structure function, which derive from nCTEQ15 and EPPS21 parametrization without accounting for the SRC effect. These results are presented in Fig.\ref{Fig3-NNN}.}

 To further utilize the universality property of nucleon pairs in practice, we rearrange Eq.(\ref{rfu}) in terms of $R_{EMC}$ as shown in Eq.(\ref{rfu-emc}), leading us to Eq.(\ref{remc}). This equation assists us in obtaining the nuclear weight function for each nucleus, incorporating the modified universal function, $R_f^u$, by focusing solely on particular characteristics of nuclei such as atomic number $Z$, mass number $A$, and SRC scale factor, $a_2$. Employing Eq.(\ref{remc}) together with Eq.(\ref{xf3A}), we can derive the nuclear structure function $xF^A_3$ for various nuclei. In Fig.\ref{Fig3-N}, we displayed the iron $xF^A_3$ structure function, derived from Eq.(\ref{xf3A}), while $R_{EMC}$ is sourced from Eq.(\ref{remc}), and $xF_3^p$ is obtained from nCTEQ15 PDFs. Employing this outcome for the $xF^A_3$ structure function, a comparison is shown in Fig.(\ref{Fig3-N}) between this result and that derived directly from the {{nCTEQ15 nuclear PDFs}} \cite{Kovarik:2015cma}, along with the available NUTeV and CCFR experimental data, demonstrating a good agreement among them. { By examining the outcomes, shown in Fig.\ref{Fig3-NNN} and Fig.\ref{Fig3-N}, the SRC effect was confirmed to improve the result for the $xF^A_3$ nuclear structure function.}

Ultimately, as an additional practical means to illustrate the benefit of the proposed method, we showed the result for the lead $xF^A_3$ structure function in Fig.\ref{Fig3}. It is  obtained  from Eq.(\ref{xf3A}) using $R_{EMC}$ in Eq.(\ref{remc}) along with $xF_3^p$, derived from nCTEQ15 PDFs. On the other hand, the result of $xF_3^A$ for the lead nucleus has also  been depicted, derived directly from nCTEQ15 nuclear PDFs. The two outcomes are consistent with each other and also with the CHROS available experimental data..

As a further research task, using the proposed method, we can establish a means to predict the $a_2$ SRC scale factor for various nuclei, considering the measured cross-section ratios for neutrino scattering relative to deuterium that solely involve the valence quark densities. To achieve accurate estimation, the calculation procedure can be cooperated with the measured nuclear cross  section that also includes the non-singlet $F_2$ structure functions.  We hope to report later on this issue as a new research activity.

\section*{Acknowledgments}
A.M. and M.K thanks Yazd university for the resources that were made available to conduct this project. H.A. would like to thank Malayer university and  School of Physics, Institute for Research in Fundamental Sciences (IPM), for providing the necessary assistance for this project.

\end{document}